\documentclass[aps,prl,preprint,showpacs,superscriptaddress]{revtex4-1}
\usepackage[pdftex]{graphicx} 
\usepackage{natbib}
\usepackage{booktabs}
\usepackage{color}

\newcommand{\T}{\mbox{1T'-WTe$_2$}}
\newcommand{\WTe}{WTe$_2$}
\newcommand{\marpes}{$\mu$-ARPES}

\begin{document}
	
\title{Microfocus laser-ARPES on encapsulated mono-, bi- and few-layer 1T'-WTe$_2$}

\author{Ir\`ene Cucchi}
\affiliation{Department of Quantum Matter Physics, University of Geneva, 24 quai Ernest Ansermet, CH-1211 Geneva, Switzerland}
\author{Ignacio Guti\'errez-Lezama}
\affiliation{Department of Quantum Matter Physics, University of Geneva, 24 quai Ernest Ansermet, CH-1211 Geneva, Switzerland}
\affiliation{Group of Applied Physics, University of Geneva, 24 Quai Ernest Ansermet, CH-1211 Geneva, Switzerland}
\author{Edoardo Cappelli}
\affiliation{Department of Quantum Matter Physics, University of Geneva, 24 quai Ernest Ansermet, CH-1211 Geneva, Switzerland}
\author{Siobhan McKeown Walker}
\affiliation{Department of Quantum Matter Physics, University of Geneva, 24 quai Ernest Ansermet, CH-1211 Geneva, Switzerland}
\author{Flavio Y. Bruno}
\affiliation{Department of Quantum Matter Physics, University of Geneva, 24 quai Ernest Ansermet, CH-1211 Geneva, Switzerland}
\author{Giulia Tenasini}
\affiliation{Department of Quantum Matter Physics, University of Geneva, 24 quai Ernest Ansermet, CH-1211 Geneva, Switzerland}
\affiliation{Group of Applied Physics, University of Geneva, 24 Quai Ernest Ansermet, CH-1211 Geneva, Switzerland}
\author{Lin Wang}
\altaffiliation[present address: ]{Key Laboratory of Flexible Electronics (KLOFE) \& Institute of Advanced Materials (IAM), Jiangsu National Synergetic Innovation Center for Advanced Materials (SICAM), Nanjing Tech University (Nanjing Tech), 30 South Puzhu Road, Nanjing 211816, China}
\affiliation{Department of Quantum Matter Physics, University of Geneva, 24 quai Ernest Ansermet, CH-1211 Geneva, Switzerland}
\affiliation{Group of Applied Physics, University of Geneva, 24 Quai Ernest Ansermet, CH-1211 Geneva, Switzerland}
\author{Nicolas Ubrig}
\affiliation{Department of Quantum Matter Physics, University of Geneva, 24 quai Ernest Ansermet, CH-1211 Geneva, Switzerland}
\affiliation{Group of Applied Physics, University of Geneva, 24 Quai Ernest Ansermet, CH-1211 Geneva, Switzerland}
\author{C\'eline Barreteau}
\altaffiliation[present address: ]{Institut de Chimie et des Mat\'eriaux Paris Est - UMR 7182, 2-8 rue H. Dunant 94320 THIAIS, France}
\affiliation
{Department of Quantum Matter Physics, University of Geneva, 24 quai Ernest Ansermet, CH-1211 Geneva, Switzerland}
\author{Enrico Giannini}
\affiliation
{Department of Quantum Matter Physics, University of Geneva, 24 quai Ernest Ansermet, CH-1211 Geneva, Switzerland}
\author{Marco Gibertini}
\affiliation{Department of Quantum Matter Physics, University of Geneva, 24 quai Ernest Ansermet, CH-1211 Geneva, Switzerland}
\affiliation{National Centre for Computational Design and Discovery of Novel Materials (MARVEL), \'Ecole Polytechnique F\'ed\'erale de Lausanne, CH-1015 Lausanne, Switzerland}
\author{Anna Tamai}
\affiliation{Department of Quantum Matter Physics, University of Geneva, 24 quai Ernest Ansermet, CH-1211 Geneva, Switzerland}
\author{Alberto F. Morpurgo}
\affiliation{Department of Quantum Matter Physics, University of Geneva, 24 quai Ernest Ansermet, CH-1211 Geneva, Switzerland}
\affiliation{Group of Applied Physics, University of Geneva, 24 Quai Ernest Ansermet, CH-1211 Geneva, Switzerland}
\author{Felix Baumberger}
\affiliation
{Department of Quantum Matter Physics, University of Geneva, 24 quai Ernest Ansermet, CH-1211 Geneva, Switzerland}
\affiliation{Swiss Light Source, Paul Scherrer Institute, CH-5232 Villigen, Switzerland}

\begin{abstract}
	Two-dimensional crystals of semimetallic van der Waals materials hold much potential for the realization of novel phases, as exemplified by the recent discoveries of a polar metal in few layer \T{} and of a quantum spin Hall state in monolayers of the same material. Understanding these phases is particularly challenging because little is known from experiment about the momentum space electronic structure of ultrathin crystals. Here, we report direct electronic structure measurements of exfoliated mono- bi- and few-layer \T{} by laser-based micro-focus angle resolved photoemission. This is achieved by encapsulating with monolayer graphene a flake of \WTe{} comprising regions of different thickness. Our data support the recent identification of a quantum spin Hall state in monolayer 1T'-WTe$_2$ and reveal strong signatures of the broken inversion symmetry in the bilayer. We finally discuss the sensitivity of encapsulated samples to contaminants following exposure to ambient atmosphere.
\end{abstract}

\maketitle

Two-dimensional (2D) van der Waals (vdW) materials host fascinating phases of quantum matter. A particularly striking example is \WTe. In the bulk, its most stable 1T' structure has a non-symmorphic space group $(Pnm2_1)$, shows giant non-saturating magnetoresistance~\cite{Ali2014a}, pressure induced superconductivity~\cite{Kang2015a, Pan2015} and has been proposed as a candidate type-II Weyl semimetal~\cite{Soluyanov2015}, although the latter has not been confirmed experimentally to date~\cite{Bruno2016,Armitage2018}. Exfoliated crystals remain metallic down to 3 Te-W-Te units when encapsulated in hexagonal boron nitride (h-BN)~\cite{Fei2017,Fatemi2017}
\footnote{The bulk primitive unit cell of \T{} contains 2 Te-W-Te layers and 4 formula units. However, the thickness of thin samples is given in the literature and throughout this paper in multiples of individual Te-W-Te{} trilayers.}. 
The monolayer (ML) is a robust 2D topological insulator with helical edge modes producing a quantum spin Hall state~\cite{Qian2014,Fei2017,Tang2017,Wu2018,Shi2018,Xu2018} and becomes superconducting when doped electrostatically~\cite{Sajadi2018,Fatemi2018}. The bilayer (BL) is a topologically trivial ferroelectric insulator at low temperature and becomes metallic above $\sim 20$~K while retaining a finite polarization~\cite{Fei2017,Fei2018}. Spontaneous electrical polarization, switchable with a gate field, was also found in trilayer samples where it coexists with metallic conductivity at low temperature~\cite{Fei2018}. 

Understanding the emergence of these properties as the thickness is reduced towards a single ML is challenging, in part because many of the powerful experimental tools developed for the study of bulk single crystals currently lack the sensitivity to be applied to a single exfoliated flake of a few micron lateral dimension. 
Angle resolved photoemission (ARPES), arguably the most direct probe of the electronic band structure, is an exception in the sense that it is, at least in principle, sensitive to a single ML. Studying 2D van der Waals materials by ARPES proved challenging though. 
One approach is the \textit{in-situ} deposition of vdW materials, which produces large samples, compatible with standard high-resolution ARPES instruments. While this has been used successfully to study several transition metal dichalcogenide (TMD) ML systems~\cite{Zhang2013,Tang2017}, the deposition of multilayer TMDs or heterostructures is often not possible. Moreover, epitaxial growth limits the number of substrates and generally results in multiple structural domains, which complicates the interpretation of momentum space resolved but real space averaging measurements~\cite{Tang2017}. Alternatively, several groups have used synchrotron based micro-focus ARPES on samples prepared by \textit{ex-situ} micromechanical exfoliation of bulk crystals~\cite{Jin2013,CoyDiaz2015,Yuan2016,Pierucci2016,Wilson2017}. However, little is known about methods to prepare suitable samples with a surface quality rivaling that of bulk single crystals cleaved in ultra-high vacuum (UHV). To date \marpes{} experiments have been limited to a few semiconducting TMDs, which are stable in air and can be annealed at high temperature to clean the surface~\cite{Jin2013,CoyDiaz2015,Yuan2016,Pierucci2016,Wilson2017}. This approach facilitates the preparation of samples but is not readily applicable to semimetallic TMDs such as \T, superconductors and density wave systems such as FeSe or NbSe$_2$ or topical vdW magnets such as CrI$_3$, which all are too reactive and/or decompose at high temperature.

Here, we report laser-based \marpes{} experiments on exfoliated flakes of the reactive semimetal \T. We show that encapsulation of a flake with ML graphene under protective atmosphere provides ARPES data of a quality comparable to that of bulk single crystals cleaved in UHV. We clearly resolve the inverted band gap underlying the quantum spin Hall state in ML \T{} and find a strong Rashba-like spin-splitting in the BL arising from its broken inversion symmetry.
Our results further demonstrate that laboratory based \marpes{} instruments are a promising alternative to micro-focus synchrotron beamlines.

\begin{figure}[tb]
	\includegraphics[width=0.7\textwidth]{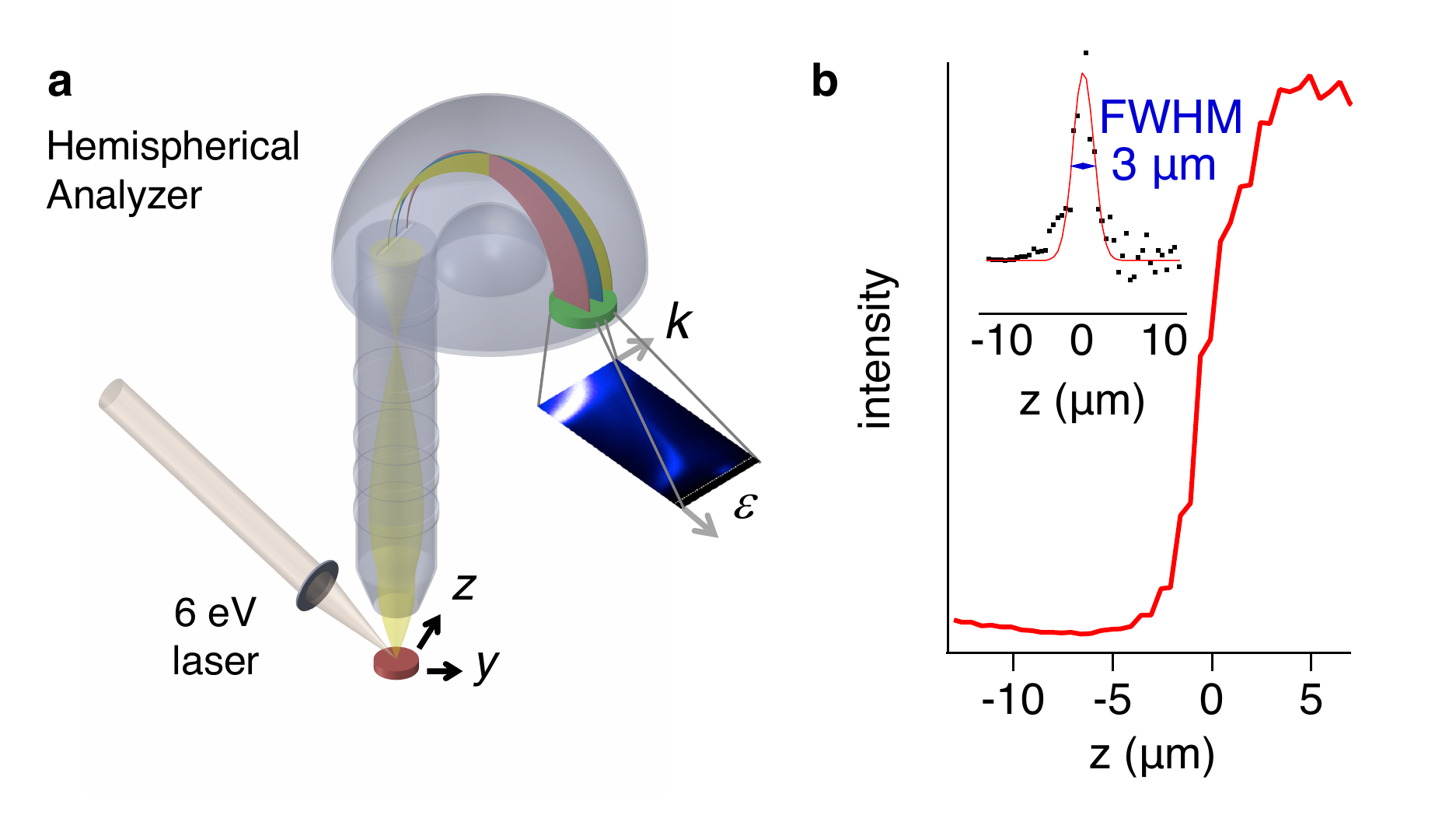}
	\caption{(a) Sketch of the \marpes{} setup. For details, see main text. (b) Knife edge scan across the sharp transition from the SiO$_2$ substrate to a Au contact. The inset shows the derivative of the line scan with a Gaussian fit indicating a full width half maximum (FWHM) of $\sim3$~$\mu$m.}
\end{figure}

The ARPES measurements reported here were performed in a custom built micro-ARPES system illustrated schematically in Fig.~1(a). A continuous wave laser source from LEOS solutions providing up to $10^{15}$~photons/s with 206~nm wavelength (6.01~eV) in a $\mu$eV bandwidth was used as excitation source. The laser beam was first expanded and then refocused using a 1" aberration corrected lens with $\sim65$~mm focal lenght mounted in UHV on a motorized 3-axes translator. Photoelectrons were collected and analyzed with an MB Scientific hemispherical analyzer equipped with a deflector lens capable of acquiring 2D $k$-space maps without rotating the sample. Typical energy and momentum resolutions were 0.003~\AA$^{-1}$ / 2~meV. Samples have been mounted on a conventional 6-axes ARPES manipulator described in Ref.~\cite{Hoesch2017}. All experiments were performed at pressures $<10^{-10}$~mbar and a manipulator temperature of $\sim 4.5$~K measured on the sample receptacle. The actual temperature of the exfoliated flake determined from the width of the Fermi cutoff was in the range of $30-40$~K. Sample positions were scanned with a stepper motor driven $xyz$-stage with 100~nm resolution and $< 1\;\mu$m bidirectional reproducibility. The effective spatial resolution of the system has been determined from line scans across Au contacts on a SiO$_2$ substrate. As shown in Fig.~1(b), this indicates a resolution of $3\;\mu$m. Drift of the sample position was found to be significant during cool down but negligible in thermal equilibrium.

\begin{figure*}[tb]
	\includegraphics[width=0.9\textwidth]{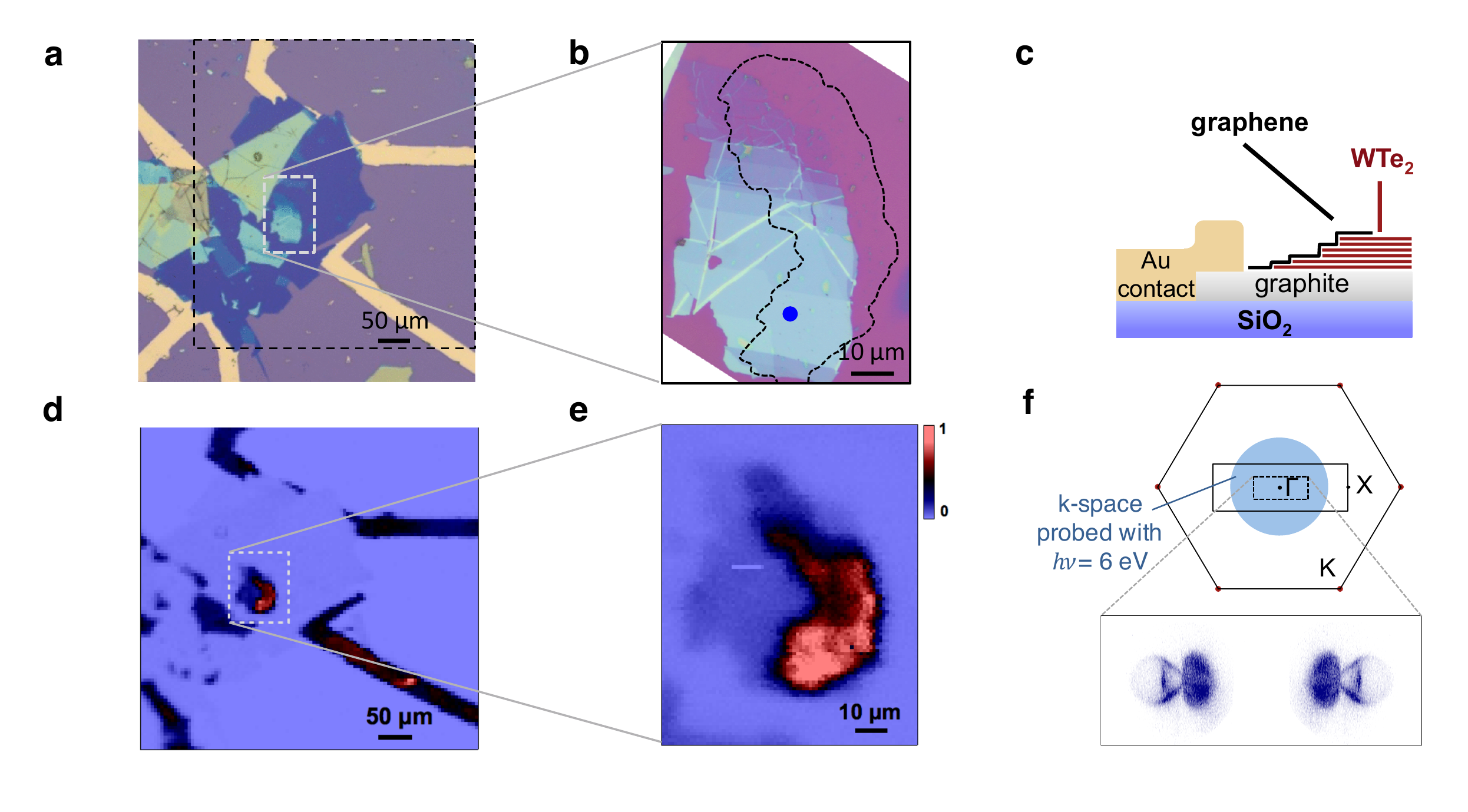}
	\caption{(a) Micrograph of the full assembly with Au contacts (yellow), graphite bottom electrode (blue) and different \WTe{} flakes (turquoise) on the SiO$_2$ substrate (violet). (b) Zoom-in showing the encapsulated flake in the area marked by a dashed white line in (a). The boundary of the graphene encapsulation layer is shown by a black dotted line. The blue circle with 3~$\mu$m diameter illustrates the spatial resolution of the \marpes{} experiments. (c) Schematic of the sample studied in this work.  (d) Scanned \marpes{} image of the area marked by the black dashed rectangle in (b). (e) Magnified \marpes{} image of the partially encapsulated \WTe{} flake (dashed grey rectangle in (d)). (f) Brillouin zone of bulk \T{} (black rectangle) and graphene (hexagon). The magnified inset shows the Fermi surface of cleaved bulk \T{} measured with $h\nu=6$~eV.}
\end{figure*}

Optical micrographs and the schematic configuration of the vdW heterostructure used for our experiments are shown in Fig.~2(a-c).
A graphite bottom electrode is placed on a SiO$_2$ wafer and contacted with Ti/Au lines defined with conventional nano-fabrication techniques  to ground the assembly. The bottom electrode provides at the same time an atomically flat surface with minimal contamination from adsorbates.
WTe$_2$ crystals were synthesized with the chemical vapor transport (CVT) method~\cite{Wang2015} and exfoliated onto a Si/SiO$_2$ substrate in the protective atmosphere of a glovebox with sub-ppm contamination of water and a concentration of O$_2$ below 20~ppm. For our experiments, we isolated the flake shown in Fig.~2(b) whose thickness determined from the optical contrast on SiO$_2$ ranges from a single ML to $\sim 12$~nm. The WTe$_2$ flake was then picked up with a graphene ML on a PC/PDMS stamp and transfered onto the bottom graphite electrode using an all-dry transfer method~\cite{Mayorov2011, Zomer2014}. This ensures that the interface between WTe$_2$ and the encapsulating graphene ML is free from PC residues. The relative orientation between the thick bottom graphite electrode, the WTe$_2$ flake and the graphene ML used to encapsulate it was not controlled. More details about the fabrication of the vdW heterostructure are provided in supplementary information.
The assembled heterostructure was transported under glovebox atmosphere to the ARPES system using a specially designed suitcase and pumped down to UHV pressures without ever exposing it to ambient air. The sample was neither annealed nor cleaned in any other way prior to the first ARPES measurements. The thickness of the relevant areas of the WTe$_2$ flake determined from the optical contrast was confirmed after the \marpes{} experiments using \textit{ex-situ} atomic force microscopy (supplementary figure S2).

Fig.~2(d) demonstrates imaging of the full assembly by laser-based \marpes. Scanning the sample while collecting photoelectrons near normal emission with energies within 0.3~eV from the Fermi level shows high contrast between the Au contact lines and the SiO$_2$ substrate. High intensity is also observed for the encapsulated part of the \WTe{} flake (marked by a black line in Fig.~2(b)) providing a first indication for the effective protection of its low-energy electronic states. The graphene encapsulation layer, on the other hand, is completely invisible in our \marpes{} images and even the thick graphite bottom electrode shows minimal intensity. This is a direct consequence of the energy and momentum space resolution of ARPES. As shown in Fig.~2(f), the Fermi surface of \T{} is well separated in $k$-space from the low-energy electronic states of graphene at the K-points. Moreover, within the momentum space accessible at 6~eV photon energy, the lowest lying graphene excitations arise from the \mbox{$\sigma$-band} located several eV below the chemical potential. Direct transitions from graphene or graphite initial states are thus not possible within the entire energy and momentum range probed in our experiments.

\begin{figure}[tb]
	\includegraphics[width=0.6\textwidth]{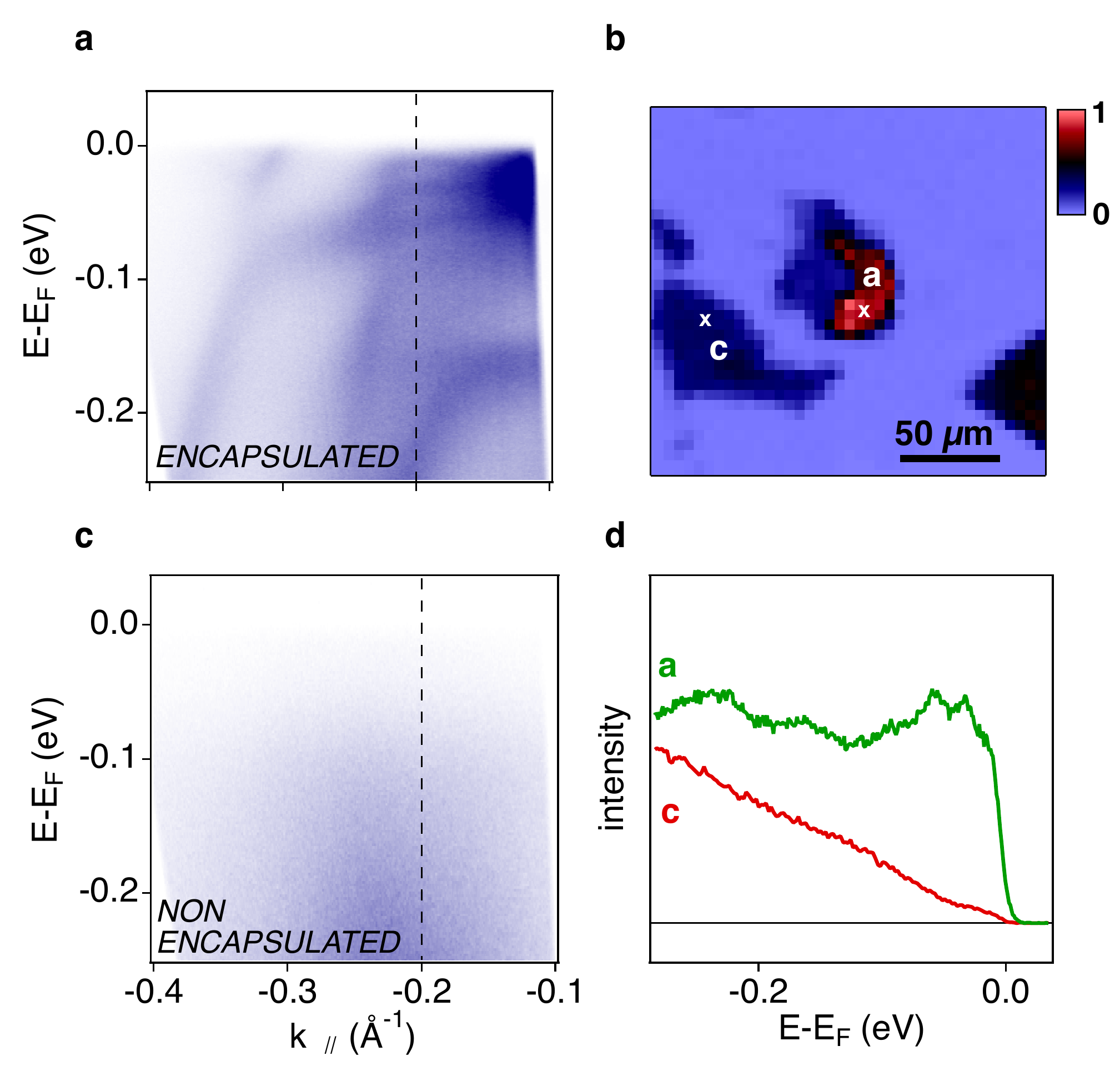}
	\caption{Effect of encapsulation on the electronic states. (a) Band dispersion of encapsulated mulitlayer \T{} along $\Gamma$X measured at the position indicated in the real space image shown in (b). (c) \marpes{} data acquired under identical conditions on a non-encapsulated flake of similar thickness at the position indicated in (b). (d) Energy distribution curves at $k_{\parallel}=0.2$~\AA$^{-1}$ (black dashed line) extracted from the data in (a), (c).}
\end{figure}

The effect of encapsulation is demonstrated directly in Fig.~3, where we compare \marpes{} data from the bulk-like part of the encapsulated \WTe{} flake with data from a different flake of similar thickness that was not encapsulated. The data taken on encapsulated \WTe{} show multiple well defined dispersive states with an overall data quality that is comparable to high-resolution ARPES data from cleaved bulk samples~\cite{Pletikosic2014,Bruno2016}. In contrast, the non-encapuslated flake shows a largely featureless ARPES spectrum with strongly reduced intensity at the chemical potential as it is typical for the heavily contaminated surface of bulk \WTe. This is remarkable, considering that both flakes were exposed to the same environment and implies
an effective protection of the encapsulated flake and/or an active self-cleansing effect during the encapsulation with graphene~\cite{Mayorov2011,Kretinin2014}.

\begin{figure*}[htp!]
	    \includegraphics[width=0.59\textwidth]{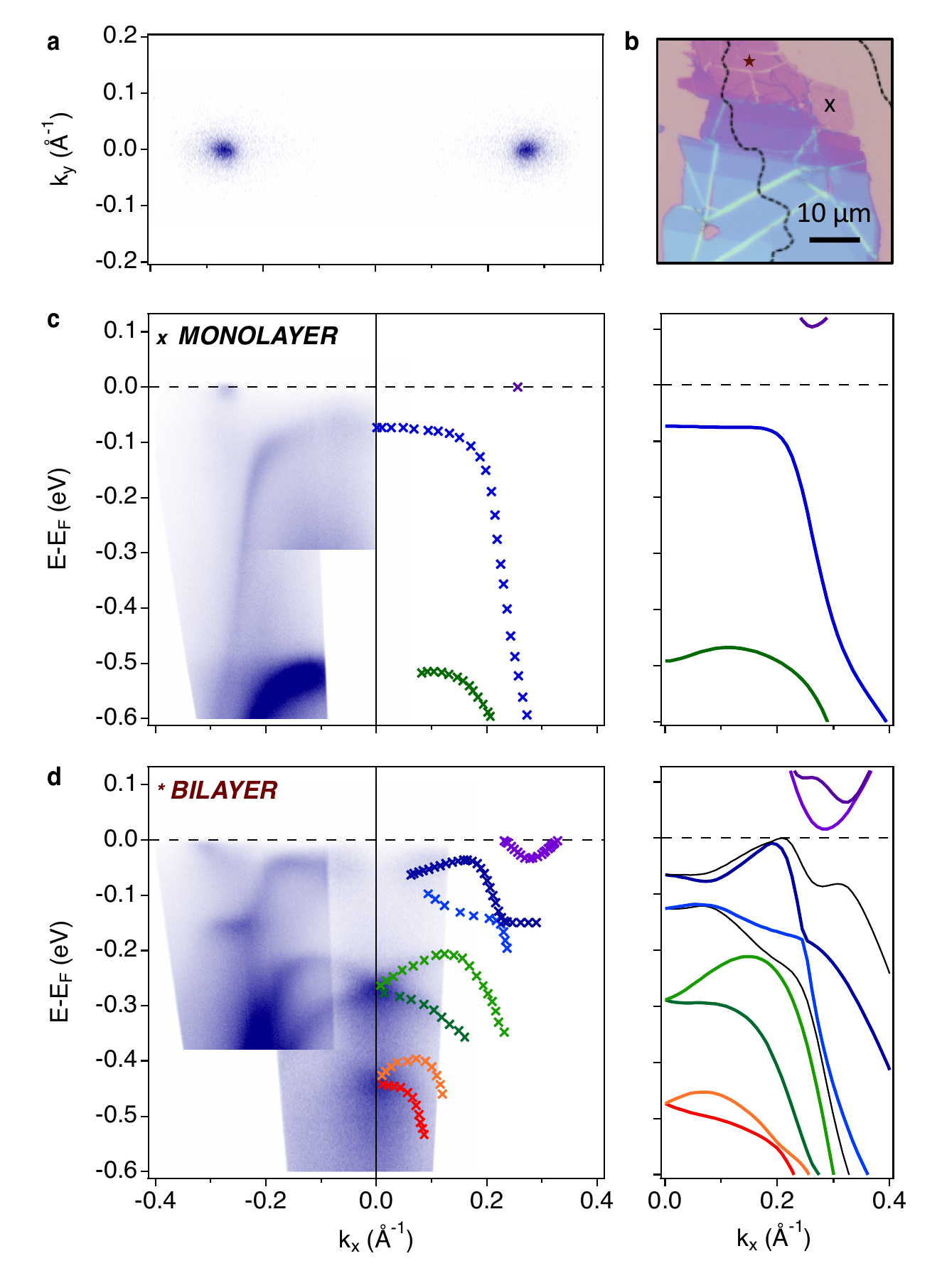}
  \caption{Electronic structure of ML and BL \T. (a) Fermi surface of ML \T. (b) Micrograph of WTe$_2$ as exfoliated on SiO$_2$. The black dotted line marks the part of the flake that has been encapsulated with monolayer graphene. (c,d) Band dispersion along $\Gamma$X taken in the ML and BL regions of the sample marked by a black cross and dark red star in (b), respectively. The abrupt breaks in contrast in the false color plots arise from combining data taken with different light polarization and in different experimental geometries. The ML was measured with $s$-polarization at high $k_\parallel$ and $p$-polarization near normal emission, while both data sets merged to obtain the BL dispersion were taken with $s$-polarization. 
Band dispersions extracted from experiment are shown with crosses and colored to facilitate comparison with the DFT band structures shown on the right. DFT calculations have been performed for isolated ML and BL with the Heyd–Scuseria–Ernzerhof hybrid functional~\cite{Heyd2003} as implemented in the VASP code~\cite{Kresse1996} (see supplementary information for more details).}
\end{figure*}

Having demonstrated well defined electronic states in encapsulated flakes, we proceed to investigate the electronic structure of ML and BL \T{} (Fig.~4).  Corresponding regions were identified in the optical micrograph, as indicated in Fig.~4(b), and measured for two different polar angles to cover an extended $k$-space range. We start by discussing the ML data. The primitive unit cell of bulk \T{} contains two \WTe{} layers. Isolating a single ML will thus profoundly change the symmetry and electronic structure. Most importantly, a ML is inversion symmetric, whereas all multilayers lack an inversion center. This causes a relatively simple electronic structure with spin-degenerate states in the ML. Our data in Fig.~4(c) show two fully occupied hole like bands and small electron pockets at $k_x\approx\pm0.3$~\AA$^{-1}$ that barely touch the chemical potential. 
From our density functional theory (DFT) calculations, we identify these states with the conduction band minimum (CBM) formed by a combination of $d_{yz}$ orbitals on the W atoms with positive overall parity $(d_{yz}^+)$, the $d_{xz}^+$ valence band maximum (VBM), and a $d_{z^{2}}^{-}$ state (with negative parity) at $\sim-0.55$~eV, in agreement with earlier theoretical work and an ARPES study on MBE grown films~\cite{Tang2017,Choe2016,Qian2014}.
This confirms the $d-d$ inversion of bands with opposite parity~\cite{Choe2016} causing the spin-momentum locked topological edge states of ML \T~\cite{Fei2017,Tang2017,Wu2018,Shi2018}. In supplementary Fig.~S4, we directly confirmed edge conduction on samples exfoliated from the same set of crystals.

From fits of energy distribution curves we determine a CBM in our ML sample of $-9(5)$~meV and a VBM of $-75(5)$~meV. We thus estimate a gap of 66(7)~meV, slightly larger than the 55~meV reported in MBE grown films~\cite{Tang2017}. Because of the very small Fermi energy in our sample we cannot directly resolve distinct Fermi crossings within a single conduction band pocket. We thus estimate an upper bound of the carrier density from the half width at half maximum of the momentum distribution curve at the chemical potential which must be greater than the Fermi wave vector $k_F$ (see supplementary Fig.~S3). 
This indicates a carrier density $n_{2D}= k_F^2/\pi < 10^{12}$~cm$^{-2}$ for the 2 nearly isotropic spin-degenerate conduction band pockets, which is below the onset of metallic conduction reported in Ref.~\cite{Fei2017}.

BL \T{} is the first atomically thin vdW material showing ferroelectric polarization. Although the origin of ferroelectricity is not fully established yet, transport evidence points to an important role of itinerant electronic states~\cite{Fei2018}. A characterization of the electronic band structure of BL \T{} is thus of particular interest. This can readily be achieved on our encapsulated sample, which has a BL region of several microns lateral dimension but would be difficult by any other means.
Remarkably, our \marpes{} data from the BL shows a surprisingly rich band structure, which cannot be approximated starting from the ML band structure by introducing a simple splitting into bonding and antibonding states. 
We attribute the additional complexity in the band structure predominantly to the strong effects of inversion symmetry breaking in BL \T, which lifts the spin degeneracy. Together with the doubling of the unit cell, one thus expects a four fold increase in the number of bands. This is indeed evident in the calculation, which shows 8 valence bands in the energy range of the two highest valence bands of the ML. The strong effect of the inversion symmetry breaking is most striking in the doublet colored in green, which shows the characteristic crossing of Rashba spin-split bands, imposed by the Kramers degeneracy at the time reversal invariant $\Gamma$ point. 
We note that not all bands predicted by the calculation are detected in experiment. This can be attributed to matrix element effects and is not unusual in ARPES.

Compared to the ML, we find a strongly increased carrier density of $n_{2D}\approx 3.6\cdot10^{12}$~cm$^{-2}$ and a vanishing, if not slightly negative gap. While the origin of these changes are not clear yet, we note that our observations are consistent with the onset of metallic conductivity above $\sim 20$~K~\cite{Fei2017} considering that the actual sample temperature in our experiments is in the range of $30-40$~K. 
We remark, however, that the size and nature of the gap in ML and BL \T{} is not yet established conclusively. DFT calculations in the generalized gradient approximation show a band overlap in both cases (see supplementary Fig.~S5). This can be rectified, as done here, by using a hybrid functional providing a more realistic description of electron correlations. 
While such calculations reproduce the overall band structure, as shown in Fig.~4, quantitative discrepancies remain. Most notably, our hybrid functional calculations overestimate the gap in ML \T{} seen in ARPES by about a factor of two.
Together with the abrupt closure of the gap at relatively low temperature reported in Ref.~\cite{Fei2017}, this suggests a non-negligible and not yet fully understood role of many-body interactions. We also point out that the gap is affected by slight changes in the crystal structure (see Fig.~S5) or by vertical electric fields, which depend on the encapsulating material and are hard to fully eliminate in experiments. Such field effects should be particularly pronounced in the BL with its strong inversion symmetry breaking and electrical polarization. 

\begin{figure*}[tb]
	\includegraphics[width=0.9\textwidth]{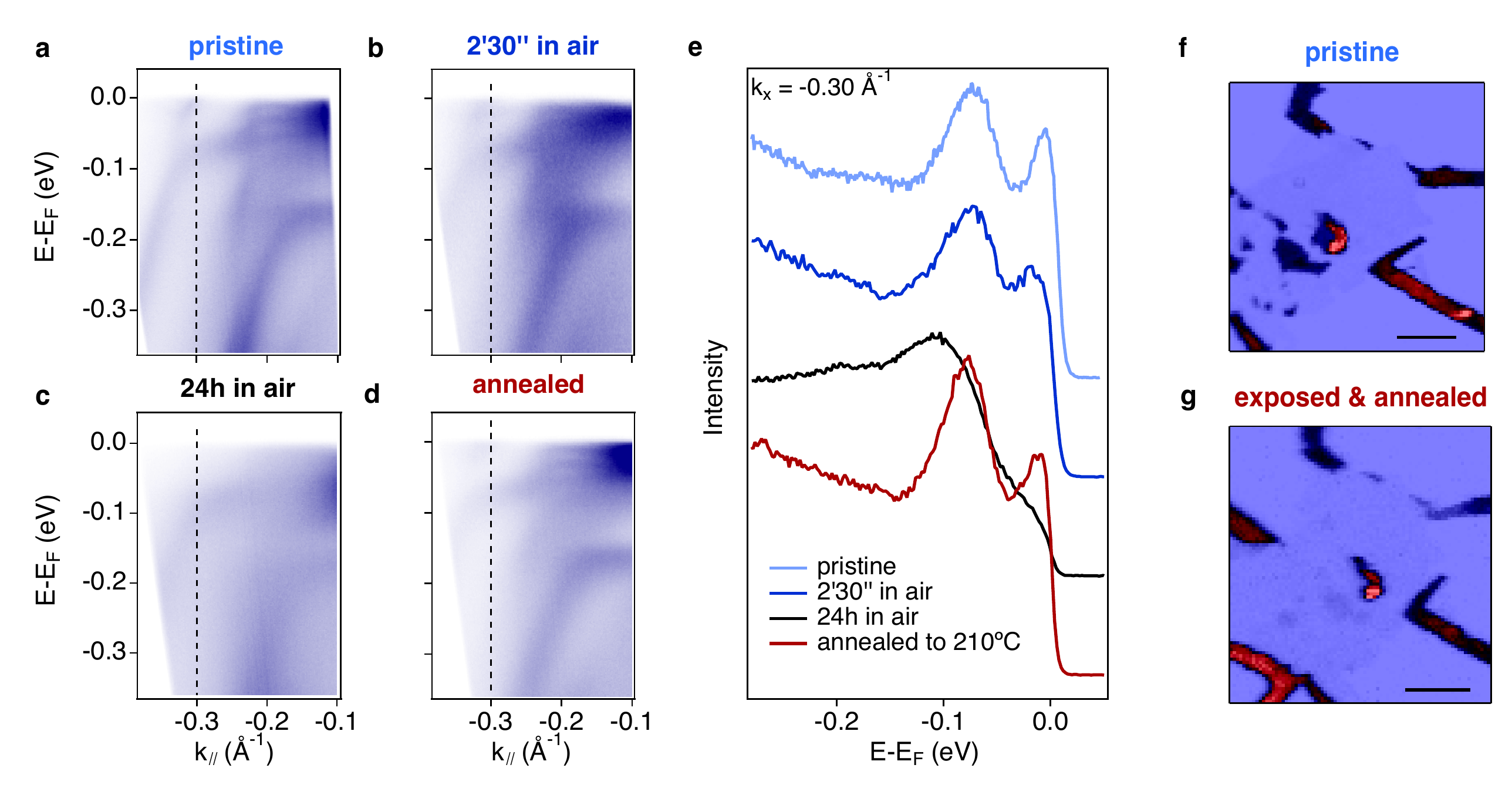}
	\caption{(a-d) \marpes{} band structure measurements of encapsulated multilayer \T, compared to data taken at the same position after exposure to air and subsequent annealing at 210$^{\circ}$~C. (e) Energy dispersion curves at $k_{\parallel}=-0.3$~\AA$^{-1}$ extracted from the data sets in (a-d). (f,g) Real-space \marpes{} images of the pristine sample and of the same area after exposure to air and annealing in UHV. The length of the scale bar is 100~$\mu$m.}
\end{figure*}

In Fig.~5, we investigate the robustness of our sample design against aging. To this end, we reproduce the \marpes{} data of bulk like \WTe{} from Fig.~3(a) and compare them with spectra taken at the same position following exposure to ambient atmosphere. The contact with air causes clear degradation in the form of a sizeable broadening of the spectra although we find remnants of the pristine band structure even after exposure to air for 24~h. 
Remarkably, the electronic structure of the encapsulated sample recovers almost completely after mild annealing in UHV (2' at $210^{\circ}$~C). This is in sharp contrast to the behavior of the exposed surface of \WTe{} reported in Ref.~\cite{Liu2017}. The irreversible and almost complete degradation of non-encapsulated \WTe{} during a cycle of exposure to air and subsequent annealing is also evident in the spatial maps of the low-energy spectral weight shown in Fig.~5(f,g).
On the pristine sample (Fig.~5(f)), we detect signal from the entire \WTe{} flake studied in \mbox{Figs.~2-5}, including its non-encapsulated part, as well as from multiple other non-encapsulated flakes, whereas at the end of the cycle, significant intensity is observed only for the Au contacts and the encapsulated part of the \WTe{} (Fig.~5(g)).

Our \marpes{} data from the pristine encapulated flake show typical linewidths of \mbox{$\sim 30$~meV}, which compares favorably to MBE grown films~\cite{Tang2017} and to \marpes{} studies of much more inert semiconducting TMDs~\cite{Jin2013,CoyDiaz2015,Yuan2016,Pierucci2016,Wilson2017}. 
We partially attribute this to the self-cleansing effect between two atomically flat surfaces when 2D materials have a high affinity reported before~\cite{Mayorov2011,Kretinin2014}. This interpretation is supported by our AFM images, which show bright white spots that are suspected to be pockets of trapped contaminants pushed away during the encapsulation to leave extended interfaces between the TMD and graphene atomically clean. 
The highly reversible behavior in \mbox{exposure / annealing} cycles further demonstrates that encapsulation with graphene largely prevents oxidation of \WTe{} observed in ambient air~\cite{Woods2017} and suggests that the predominant broadening mechanism following exposure to air is a long-range chemical potential variation caused by charged, weakly-bound contaminants on the surface of graphene, which can be desorbed in UHV by mild annealing.
We finally note that the data quality obtained on bulk-like encapsulated \WTe{} showed no signs of degradation over several months in UHV including more than a dozen temperature cycles between liquid helium and room temperature. This is in strong contrast to typical lifetimes of cleaved bulk samples for conventional ARPES experiments of 12-24 hours.

In conclusion, we have demonstrated that encapsulation of \T{} with graphene is suitable to obtain high-quality electronic structure data using \marpes. This opens the way for electronic structure measurements on a broad range of previously inaccessible ultrathin TMDs including 2D magnets, charge density wave systems and superconductors such as FeSe or NbSe$_2$, on exfoliated Dirac and Weyl semimetals, or on reactive 2D semiconductors such as phosphorene.
In addition, we showed that the use of deep UV lasers  is a promising alternative to synchrotron-based \marpes. Together, these advances should promote a more widespread use of ARPES in the study of 2D TMDs and heterostructures thereof.


\begin{acknowledgements}

We gratefully acknowledge discussions with David Cobden, Hugo Henck, Simone Lisi and Sara Ricc\`o. 
This work was supported by the Swiss National Science Foundation (SNSF) Div. II and the SNSF Synergia program. MG and NU were supported by SNSF Ambizione fellowships. Simulation time was provided by CSCS on Piz Daint (project id s825).

\end{acknowledgements}

\bibliography{exf_WTe2_02}

\end{document}